\documentstyle[psfig]{l-aa}
\def \etal     {et al.}
\def \ie       {i.\,e.}
\def \eg       {e.\,g.}
\def \vLSR     {\hbox{${v_{\rm LSR}}$}}
\def \delv     {\hbox{$\Delta v_{1/2}$}}
\def \TMB      {\hbox{$T_{\rm MB}$}}
\def \Tkin     {\hbox{$T_{\rm kin}$}}
\def \Tsys     {\hbox{$T_{\rm sys}$}}
\def \HII      {H\,{\sc ii}}
\def \kms      {\hbox{${\rm km\,s}^{-1}$}}                % km/s
\def \Kkms     {\hbox{${\rm K\,km\,s}^{-1}$}}             % K*km/s
\def \percc    {\hbox{${\rm cm}^{-3}$}}                   % cm-3
\def \ccpers   {\hbox{${\rm cm}^3{\rm s}^{-1}$}}          % cm3/s
\def \hour     {\hbox{$^{\rm h}$}}                        % h
\def \minute   {\hbox{$^{\rm m}$}}                        % m
\def \arcdeg   {\hbox{$^{\circ}$}}                        % d
\def \ffs      {\hbox{$\,.\!\!^{\rm s}$}}                 % .s
\def \RA#1     {\hbox{$\alpha_{#1}$}}                     % R.A.
\def \Dec#1    {\hbox{$\delta_{#1}$}}                     % decl.
\def \MOLH     {\hbox{H$_2$}}                             % H2
\def \HthCN    {\hbox{H$^{13}$CN}}                        % H13CN
\def \HCOp     {\hbox{HCO$^+$}}                           % HCO+
\def \DCOp     {\hbox{DCO$^+$}}                           % DCO+
\def \HthCOp   {\hbox{H$^{13}$CO$^+$}}                    % H13CO+
\def \HHHp     {\hbox{H$_3^+$}}                           % H3+
\def \HHDp     {\hbox{H$_2$D$^+$}}                        % H2D+
\def \C#1      {\hbox{$^{#1}$C}}                          % C isotopes
\def \ISOC     {\hbox{$^{12}$C$/^{13}$C}}                 % 12C/13C
\newcommand{\see}[1]{$^{\rm #1)}$}
\def \bra#1    {$\left\{\makebox{\rule[-#1ex]{0pt}{#1ex}}\right.$}
\def \uspace#1 {\makebox{\rule[#1ex]{0pt}{2ex}}}
\def \dspace   {\makebox{\rule[-2ex]{0pt}{2ex}}}
\def \PASA     {{\rm Proc. Astron. Soc. Aust.}}

\begin{document}

\thesaurus{11 (09.01.1; 09.13.2; 11.01.1; 11.13.1; 12.03.3; 13.19.3) }

\title{Molecular abundances in the Magellanic Clouds}
\subtitle{II. Deuterated species in the LMC
          \thanks {Based on observations with the Swedish-ESO
          Submillimeter Telescope (SEST) at the European Southern
          Observatory (ESO), La Silla, Chile } }

\author{Y.-N.~Chin\inst{1,2}, C.~Henkel\inst{3}, T.J.~Millar\inst{4},
        J.B.~Whiteoak\inst{5} \and R.~Mauersberger\inst{6}}

\offprints{Y.-N.~Chin, ASIAA, Taiwan, einmann@biaa21.biaa.sinica.edu.tw}

\institute{
   Institute of Astronomy and Astrophysics, Academia Sinica,
   P.O.Box 1-87, Nankang, 115 Taipei, Taiwan
\and
   Radioastronomisches Institut der Universit\"at Bonn,
   Auf dem H\"ugel 71, D-53121 Bonn, Germany
\and
   Max-Planck-Institut f\"ur Radioastronomie,
   Auf dem H\"ugel 69, D-53121 Bonn, Germany
\and
   Department of Physics, UMIST,
   P O Box 88, Manchester M60 1QD, United Kingdom
\and
   Paul Wild Observatory, Australia Telescope National Facility, CSIRO,
   Locked Bag 194, Narrabri NSW 2390, Australia
\and
   Steward Observatory, The University of Arizona,
   Tucson, AZ 85721, U.S.A.
}

\date{Received date ; accepted date}

\maketitle

\begin{abstract}

   The first definite discoveries of extragalactic deuterium are reported.
   \DCOp\ has been detected in three and DCN has been measured
   in one star-forming region of the Large Magellanic Cloud (LMC).
   While the \HCOp/\DCOp\ abundance ratios are found to be 19 $\pm$ 3,
   24 $\pm$ 4, and 67 $\pm$ 18 for N113, N44BC and N159HW, respectively,
   a HCN/DCN abundance ratio of 23 $\pm$ 5 is obtained for N113.
   These results are consistent with a gas temperature of
   about 20\,K and a D/H ratio of about $1.5 \times 10^{-5}$,
   consistent with that observed in the Galaxy.
   If the cloud temperature is closer to 30\,K,
   then a D/H ratio is required to be up to an order of magnitude larger.
   Because this ratio provides a lower limit to the primordial D/H ratio, it
   indicates that the baryon mass density alone is unable to close the universe.


\end{abstract}

\section{Introduction}

   The primordial ratio of the number of deuterium to hydrogen nuclei (D/H)
   created in big bang nucleosynthesis is the most sensitive measure of
   the cosmological baryon-to-photon ratio, $\eta$, and the
   cosmological density of baryons, $\Omega_b$.
   In the interstellar medium of our Galaxy D/H = $1.6 \pm 0.3 \times 10^{-5}$
   (Linsky \etal\ 1995), which places a strict lower limit on the primordial
   abundance, because stars reduce the proportion of D in the ISM.
   On the other hand, quasar absorption line systems should
   provide suitable targets to determine primordial D abundances
   because they can sample metal-poor gas at early epochs
   where the destruction of D should be negligible.
   Carswell \etal\ (1994) and Songaila \etal\ (1994) obtained
   a high D/H $\approx$ $2.5 \times 10^{-4}$ ratio from a possible
   deuterium absorption feature toward the quasar Q0014+813.
   More recent observations by Rugers \& Hogan (1996) and Tytler \etal\ (1996)
   do not provide clear evidence in favor or against high D/H ratios at
   large redshifts and it is still difficult to rule out entirely
   the possibility that the D feature is contaminated by H having
   a velocity different from that of the principal hydrogen lines.
   Therefore, a direct measurement of atomic deuterium toward
   high redshift quasars is a potentially powerful,
   though not yet fully explored means of studying D/H.

   Irrespective of the validity of a high or low cosmological D/H value,
   the ratio appears to be small enough to preclude any easy detection
   of deuterated molecular species in interstellar space.
   Nevertheless, the zero-point energy difference between hydrogen- and
   deuterium-bearing molecules means that deuterated species
   can become heavily fractionated in cool environments.
   Models of fractionation (Brown \& Rice 1986; Millar \etal\ 1989) indicate
   that the underlying D/H ratio is consistent with that suggested by
   Linsky \etal\ (1995), implying that the baryon mass density is not
   large enough to close the universe.
   Hence measurements of deuterated species have become an important tool
   to study the physical properties and the chemical and dynamical history
   of Galactic molecular clouds (\eg\ Wootten 1987; Millar \etal\ 1989).
   In a previous study Mauersberger \etal\ (1995) presented sensitive upper
   limits to the DCN/HCN ratio in NGC\,253 and IC\,342, which are consistent
   with the warm, metal rich environment of this starburst nucleus.

\begin{figure*}
   \vspace*{-23 mm}
   \hspace*{-10 mm} \psfig{figure=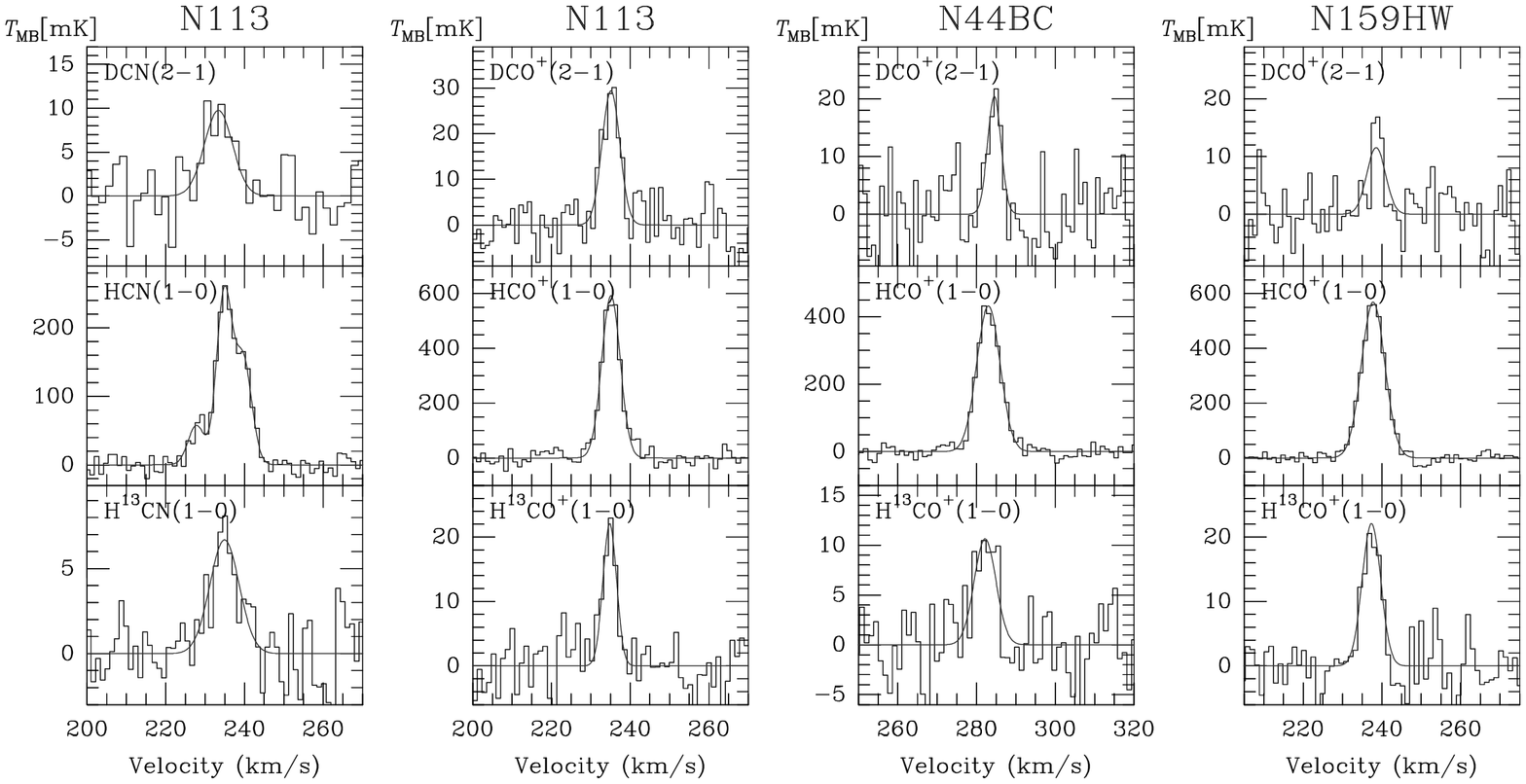,width=18 cm,angle=-90}
   \vspace*{-20 mm}
   \caption[]
           {The observed spectra of selected hydrogen-, deuterium-,
            and \C13 -bearing molecules toward
            N113 (\RA1950 \ $=$ 5\hour 13\minute 38\ffs 7,
                 \Dec1950 \ $=$ $-$69\arcdeg 25\arcmin 57\arcsec),
            N44BC (\RA1950 \ $=$ 5\hour 22\minute 10\ffs 6,
                  \Dec1950 \ $=$ $-$68\arcdeg 00\arcmin 32\arcsec), and
            N159HW (\RA1950 \ $=$ 5\hour 40\minute 04\ffs 4,
                   \Dec1950 \ $=$ $-$69\arcdeg 46\arcmin 54\arcsec)
            in the LMC}
 \label{fig:MC2-spectra}
\end{figure*}

\section{Observations}

   The data were taken in January and March 1996, using the 15-m Swedish-ESO
   Submillimetre Telescope (SEST) at La Silla, Chile.
   Two SIS receivers, one at $\lambda$\ = 3 and one at 2\,mm,
   yielded overall system temperatures, including sky noise,
   of order \Tsys\ = 250\,K on a main beam brightness temperature (\TMB) scale.
   The backend was an acousto-optical spectrometer which was split into 2
   $\times$ 1000 contiguous channels for simultaneous 3 and 2\,mm observations.
   The channel separation of 43\,kHz corresponds to
   0.09 and 0.15\,\kms\ at 145 and 88\,GHz, respectively.
   The corresponding antenna beamwidths were 35\arcsec\ and 55\arcsec\ at the
   observed line frequencies taken from Lovas (1992).

   The observations were carried out in a dual beam-switching mode (switching
   frequency 6\,Hz) with a beam throw of 11\arcmin 40\arcsec\ in azimuth.
   The on-source integration time of each spectrum varied from
   8 minutes for \HCOp\ to 18 hours for \HthCN\ toward N113.
   All spectral intensities obtained were converted to a \TMB\ scale,
   correcting for main beam efficiencies of 0.74 at 85--100 and 0.67 at
   130--150\,GHz (Dr.~L.B.G.~Knee, priv.\ comm.).
   The pointing accuracy, obtained from measurements of the SiO maser sources
   R\,Dor and U\,Men, was better than 10\arcsec.

\begin{table*}
   \caption[]
           {Parameters of the Observed Molecular Lines}
 \label{tbl:MC2-parameter}
   \begin{flushleft}
   \begin{tabular}{l l r@{.}l r@{}l r c r@{.}l r@{.}l@{~$\pm$~}l}
   \hline
   \multicolumn{2}{l}{Molecule}
         & \multicolumn{2}{c}{\hspace{-3.2mm} Frequency}
         & \multicolumn{2}{l}{\TMB}   & r.m.s.   & \vLSR  \uspace1
         & \multicolumn{2}{c}{\delv}  & \multicolumn{3}{c}{$\int$ \TMB\,d$v$} \\
       % & Velocity Range
%
   \multicolumn{2}{l}{\& Transition}
         & \multicolumn{2}{c}{\hspace{-3.2mm} [GHz]}
         & \multicolumn{2}{l}{[mK]}   & [mK]     & [\kms] \dspace
         & \multicolumn{2}{c}{[\kms]} & \multicolumn{3}{c}{[\Kkms]}           \\
       % & [\kms]
%
   \hline
\uspace1
   {\bf N113} \\
   DCN \see{a}    & $J$=2--1     & 144&077321
                  &  10&   &   8 &  233.4 & 8&4 & 0&086 & 0.011 \\% & (224,247)
   HCN \see{b}    & $J$=1--0 \bra2 \begin{tabular}{l}
                    $F$=1--1  \\ $F$=2--1  \\ $F$=0--1  \end{tabular}
                  & \multicolumn{2}{c}{\hspace{0.5mm} \begin{tabular}{r@{.}l}
                                    88&630416   \hspace{-1mm}\\
                                    88&631847   \\   88&633936 \end{tabular}}
                  & \begin{tabular}{r}  153  \\  247  \\   57  \end{tabular}
                    \hspace*{-3.4mm} &
                           &  19 &  234.9 & 4&7 & 2&33  & 0.04  \\% & (224,247)
   \HthCN\ \see{a}& $J$=1--0     &  86&340184
                  &   6&.7 &   3 &  235.0 & 8&8 & 0&065 & 0.006 \\% & (224,247)
\uspace1
   \DCOp          & $J$=2--1     & 144&828000
                  &  29&   &  10 &  235.1 & 5&3 & 0&166 & 0.011 \\% & (228,242)
   \HCOp          & $J$=1--0     &  89&188518
                  & 593&   &  39 &  235.1 & 5&5 & 3&56  & 0.06  \\% & (228,242)
   \HthCOp        & $J$=1--0     &  86&754294
                  &  22&   &   6 &  234.8 & 4&1 & 0&105 & 0.008 \\% & (228,242)
\uspace1
   {\bf N44BC} \\
   \DCOp          & $J$=2--1     & 144&828000
                  &  20&   &  12 &  284.5 & 4&0 & 0&084 & 0.014 \\% & (275,291)
   \HCOp          & $J$=1--0     &  89&188518
                  & 432&   &  25 &  283.0 & 7&0 & 3&16  & 0.04  \\% & (275,291)
   \HthCOp        & $J$=1--0     &  86&754294
                  &  11&   &   6 &  282.2 & 6&5 & 0&067 & 0.009 \\% & (275,291)
\uspace1
   {\bf N159HW} \\
   \DCOp          & $J$=2--1     & 144&828000
                  &  12&   &   8 &  238.5 & ~~~5&6 \see{c}
                                                & 0&054 & 0.010 \\% & (228,247)
   \HCOp          & $J$=1--0     &  89&188518
                  & 569&   &  23 &  237.8 & 7&0 & 4&29  & 0.04  \\% & (228,247)
\dspace
   \HthCOp        & $J$=1--0     &  86&754294
                  &  22&   &   5 &  237.4 & 5&6 & 0&122 & 0.009 \\% & (228,247)
   \hline
   \end{tabular}
   \end{flushleft}
  {\footnotesize \begin{enumerate} \renewcommand{\labelenumi}{\alph{enumi})}
   \item The hyperfine components of DCN and \HthCN\ line cannot be resolved.
   \item The three HCN hyperfine transitions ($F$=1--1, $F$=2--1, $F$=0--1)
         have been resolved by a Gaussian fit.
         While \TMB\ values for each component are given, the total integrated
         line intensity refers to the entire line.
   \item During the Gaussian fit value of \delv\ has been fixed by using that
         determined from the \HthCOp(1--0) transition.
  \end{enumerate} }
\end{table*}

\section{Results}

   Toward the prominent star-forming region N113 we have detected \HCOp, HCN,
   and their rare isotopic species \HthCOp, \DCOp, \HthCN, and DCN\@.
   \HCOp, \HthCOp, and \DCOp are also observed toward
   N159HW (position defined by Hunt \& Whiteoak 1994) and N44BC.
   The spectra and line parameters obtained from Gaussian fits are displayed in
   Fig.\,\ref{fig:MC2-spectra} and Table\,\ref{tbl:MC2-parameter}.

   Assuming optically thin line emission,
   we can derive from the line intensity ratios of the \C13 - and
   deuterium-bearing species the ratio of column densities.
   We find
\begin{displaymath}
   \frac{N_1}{N_2} = \alpha \frac{B_2}{B_1}
            \frac{g_{{\rm l}_{1}}}{g_{{\rm u}_{1}}}
            \frac{g_{{\rm u}_2}}{g_{{\rm l}_2}}
            \frac{(1 - e^{-h\nu_2/kT_2})}{(1 - e^{-h\nu_1/kT_1})}
            \frac{(J_{\nu_2,T_2} - J_{\nu_2,2.7})}
                 {(J_{\nu_1,T_1} - J_{\nu_1,2.7})}\
            \frac{I_1}{I_2},
\end{displaymath}
\begin{equation}
   \alpha = \frac{e^{2\,hB_1/kT_1}}{3}      \hbox{\hspace{7mm}}
   {\rm and}                              \hbox{\hspace{7mm}}
   J_{\nu,T} = \frac{h\nu}{k} \left( e^{h\nu/kT} - 1 \right)^{-1}
\end{equation}
   ($N$: column density; $B$: rotational constant;
   $g_l$ and $g_u$: statistical weights of lower and upper states;
   $T$: excitation temperature; $I$: integrated line intensity;
   indices 1 and 2 refer to the deuterium-bearing and to
   the \C13 -bearing or main isotopic species, respectively).
   Both \HCOp\ and HCN have large electric dipole moments (of order 3\,Debye)
   and are excited well beyond 2.7\,K only at high densities.
   At $n\,({\rm H_2}) = 10^5$ \percc\, radiative transfer calculations
   with a Large Velocity Gradient (LVG) model, spherical cloud
   geometry, and collision cross sections from Green \& Thaddeus (1974)
   yield excitation temperatures of order 4\,K for $T_{\rm kin}$ = 20--40\,K
   in the optically thin limit, both for the $J$=1--0 transition of
   \HthCN\ and for the 2--1 transition of DCN.
   These results also hold for \HCOp.
   While smaller densities would lead to excitation temperatures almost
   undistinguishable from those of the microwave background (and thus to
   negligible emission), gas at even higher densities is likely not common
   enough to contribute significantly to the observed emission.
   We thus find, neglecting differences in beam sizes, for \HCOp\
\begin{equation}
   \frac{N({\rm DCO^+})}{N({\rm H^{13}CO^+})} \approx\
            1.68 \frac{I({\rm DCO^+})}{I({\rm H^{13}CO^+})}
\end{equation}
   and for HCN
\begin{equation}
   \frac{N({\rm DCN})}{N({\rm H^{13}CN})} \approx\
            1.67 \frac{I({\rm DCN})}{I({\rm H^{13}CN})}.
\end{equation}
   With \ISOC\ $\approx$ 50 (Johansson \etal\ 1994; this is consistent
   with the line intensity ratios dereived from Table\,\ref{tbl:MC2-parameter})
   we then obtain $N$(\HCOp)/$N$(\DCOp) $\approx$ 19 $\pm$ 3 and
   $N$(HCN)/$N$(DCN) $\approx$ 23 $\pm$ 5 for N113,
   while $N$(\HCOp)/$N$(\DCOp) $\approx$ 24 $\pm$ 4 and 67 $\pm$ 18
   for N44BC and N159HW, respectively.

\begin{table*}
   \caption[]
           {Calculated abundance ratios}
 \label{tbl:MC2-model}
   \begin{flushleft}
   \begin{tabular}{c c@{~~}c@{~~}c c c@{~~}c@{~~}c c c@{~~}c@{~~}c}
   \hline
   \uspace1
   Temperature &  \multicolumn{3}{c}{10\,K}   & &  \multicolumn{3}{c}{20\,K}
             & &  \multicolumn{3}{c}{30\,K}   \\
   \cline{2-4}
   \cline{6-8}
   \cline{10-12}
   \noalign{\smallskip}
   Time (yr)   & $R$(\HHHp) & $R$(\HCOp) & $R$(HCN)
             & & $R$(\HHHp) & $R$(\HCOp) & $R$(HCN)
             & & $R$(\HHHp) & $R$(\HCOp) & $R$(HCN) \\
   \noalign{\smallskip}
   \hline
   \uspace0.5
 $10^5$           & 3.8 & 12.8 & 24.9 && 8.3 & 23.7 & 61.1 && 167 & 309 & 246 \\
   \uspace0.3
 $3.2\times 10^5$ & 3.7 & 10.3 & 14.9 && 8.2 & 17.7 & 32.5 && 166 & 257 & 253 \\
   \uspace0.3 \uspace-0.5
 $10^7$           & 4.3 & ~9.3 & 20.0 && 8.6 & 14.8 & 37.5 && 166 & 251 & 476 \\
   \hline
   \end{tabular}
   \end{flushleft}
  {\footnotesize \begin{enumerate} \renewcommand{\labelenumi}{\alph{enumi})}
   \item The ratio $R$(HX) is defined as the abundance ratio of HX to DX,
         and is tabulated for HX $=$ \HHHp, \HCOp\ and HCN
  \end{enumerate} }
\end{table*}

\section{Discussion}

   The detection of \DCOp\ and DCN in the LMC is a useful tool to study
   D/H on a larger spatial scale and in a less `chemically' processed
   environment than that provided by the Galaxy.
   Gensheimer \etal\ (1996) observed a high deuteration of water toward
   a number of Galactic hot cores despite a high kinetic temperature.
   While this can be explained in terms of a recent evaporation of
   grain mantles, this effect probably plays no role toward our
   LMC sample since we are looking on a much larger region
   which may much better reflect the average conditions of the interstellar gas
   than the highly biassed sample by Gensheimer \etal\ (1996).
   Due to the lower metallicity in the LMC, we expect grain surface reactions
   to be of lesser importance than for the Galactic interstellar medium.
   The process of chemical fractionation has, however,
   to be accounted for quantitatively.
   To this end, we have re-visited earlier models of deuterium chemistry
   (Millar \etal\ 1989; Howe \& Millar 1992) with conditions appropriate
   for clouds in the LMC (Paper I) and with
   an elemental D to H ratio of $1.5 \times 10^{-5}$.
   We have updated the chemistry to include the fast dissociative
   recombination of \HHHp\ (and \HHDp) and more accurate rate coefficients
   for neutral-neutral reactions.
   Although the general trends of fractionation can be understood,
   especially for simple species like \HCOp\, detailed models are necessary
   for quantitative comparison with the observations (Millar \etal\ 1989).

   Because deuterium fractionation depends on small differences
   in zero-point energies, the process is sensitive to
   the kinetic temperature, \Tkin, of the molecular gas.
   Multi-transition studies of CO in N113 have not been made so that
   \Tkin\ is unknown, but a value of around 20\,K is expected
   since the cloud is associated with an \HII\ region (for detailed
   modelling, see Lequeux \etal\ 1994).
   To cover the range of possibilities, we present the results of
   time-dependent chemical kinetic calculations at 10, 20 and 30\,K
   for three model times in Table\,\ref{tbl:MC2-model}.

   At low temperatures there are two main routes to D fractionation:
\begin{eqnarray}
   \HHHp  + {\rm HD} & \longleftrightarrow & \HHDp  + \MOLH \\
   \HHHp  + {\rm D} & \longrightarrow & \HHDp  +  {\rm H}
\end{eqnarray}
   where we have calculated the rate coefficient, $k_{-4}$,
   for the back reaction of \HHDp\ with \MOLH\
   using the data presented by Sidhu \etal\ (1992).
   This rate coefficient is very sensitive to temperature and has values of
   $2.1 \times 10^{-18}, 2.4 \times 10^{-13}$ and $8.3 \times 10^{-12}$
   \ccpers\ at 10, 20 and 30\,K, respectively.
   For $T \ga$ 25\,K, the back reaction is fast and
   $R$(\HHHp) is large, $\approx$ 165 at 30\,K.
   \DCOp\ is formed through reactions involving \HHDp and D atoms,
   with these species contributing 64\%\,(36\%) at 10\,K, 54\%\,(46\%) at 20\,K
   and 44\%\,(56\%) at 30\,K, to the formation rate of \DCOp.

   Our results show that the deuterium fractionation is larger than
   that observed in N113 for \HCOp\ and HCN for $T \leq$ 20\,K and
   smaller for $T$ $=$ 30\,K; the observed result that
   fractionation is larger in \DCOp\ than DCN is reproduced.

   Our results are consistent with $T \sim$ 20\,K for a D/H ratio of
   1.5 $\times$ 10$^{-5}$.
   However, the temperatures of the LMC clouds
   are not, as yet, well constrained.
   The observed abundance ratios could be reproduced in warmer
   clouds if the D/H ratio is larger than our adopted value.
   At a cloud temperature of 30\,K, one would need a D/H ratio about 10 times
   larger, \ie\ on the order of $1.5 \times 10^{-4}$.
   If there temperature is 20--30\,K as suggested by Lequeux \etal\ (1994)
   for the cloud cores (for HCN, see also Chin \etal\ 1996),
   then it appears that the D/H ratio in the LMC is similar to or larger
   than that in our Galaxy.
   This would be consistent with the LMC gas being
   less processed through stars than that in our Galaxy.
   Thus, the evidence from the LMC is consistent with an open universe as
   long as the cosmological constant is small.

\begin{acknowledgements}
   We like to thank Drs.\ J.L.\ Linsky and T.L.\ Wilson for useful comments.
   TJM is supported by a grant from PPARC.
   RM was supported by a Heisenberg fellowship
   of Deutsche Forschungsgemeinschaft.
\end{acknowledgements}

\end{document}